\documentclass[aps, prl, showpacs, twocolumn, superscriptaddress]{revtex4}

\usepackage{graphicx,color}

\begin{document}

\title{Effective Rheology of Bubbles Moving in a Capillary Tube}
\author{Santanu Sinha}
\email{Santanu.Sinha@ntnu.no}
\author{Alex Hansen}
\email{Alex.Hansen@ntnu.no}
\affiliation{Department of Physics, Norwegian University of 
Science and Technology, N-7491 Trondheim, Norway}
\author{Dick Bedeaux}
\email{Dick.Bedeaux@ntnu.no}
\author{Signe Kjelstrup}
\email{Signe.Kjelstrup@ntnu.no}
\affiliation{Department of Chemistry, Norwegian University of Science 
and Technology, N-7491 Trondheim, Norway}
\date{\today }

\begin{abstract}
We calculate the average volumetric flux versus pressure drop of
bubbles moving in a single capillary tube with varying diameter, finding
a square-root relation from mapping the flow equations onto that of a driven
overdamped pendulum.  The calculation is based on a derivation of
the equation of motion of a bubble train from considering the capillary
forces and the entropy production associated with the viscous flow.  We
also calculate the configurational probability of the positions of the bubbles.
\end{abstract}

\maketitle


Multiphase flow in porous media plays a pivotal role in a vast range of
applications in different fields such as oil recovery, soil mechanics and
hydrology \cite{d79,b88,a92,s95}.  In spite of its relevance in these
important fields, fundamental questions still linger on.  In particular,
this is true in connection with {\it steady-state\/} multiphase flow
\cite{gr93,ap95,ap99,kah02,kh02,tkrlmtf09,tlkrfm09}, which sets in after
the initial instabilities such as viscous fingering are over in e.g.\
flooding experiments.  A way to study this flow in the laboratory,
is to simultaneously inject the two immiscible fluids into the porous
medium and let them mix until a steady state where clusters and bubbles
break up and merge but in such a way that their averages remain constant
\cite{tkrlmtf09,tlkrfm09}.

Recently, the relation between average volumetric flow and excess pressure
drop across the system has been investigated both experimentally and 
theoretically \cite{tkrlmtf09,tlkrfm09,rcs11,sh12}.  The conclusion 
from these studies is that the volumetric flux depends {\it quadratically\/}
on the excess pressure.  This is in contrast to the assumptions of 
linearity commonly made when considering such systems, e.g.\ in connection 
with invoking the concepts of relative permeability in reservoir simulations
at the flow rates where capillary and viscous forces compete \cite{c06}.

One aim of this work is to derive the volumetric flux versus excess pressure
drop for a single capillary tube.  We find that there is a {\it square
root\/} singularity in this relation.  This is in contrast to a the
situation for {\it network of pores,\/} i.e., a porous medium.  We base this
calculation on the equation of motion of a bubble train in a long 
capillary tube with varying diameter, the Washburn equation \cite{w21}.  We
derive this equation following a different route than the now 91 year
old original derivation.  Lastly, we derive the probability distribution
of the configuration of bubbles in the tube.  This makes it possible to
calculate the average of any quantity associated with the flow.

\begin{figure}
\includegraphics[scale=0.5,clip]{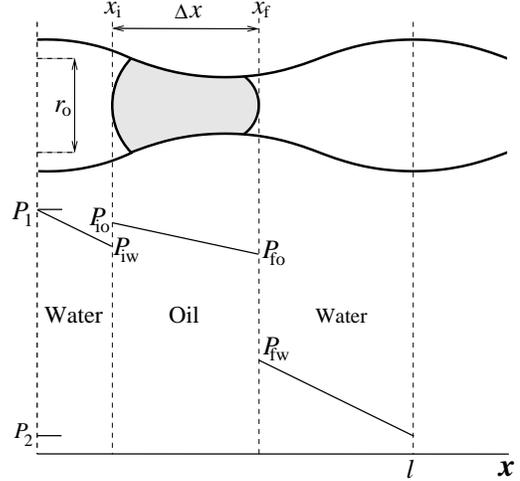}
\caption{\label{fig1} Shape of tube and corresponding pressure drops.}
\end{figure}

We assume a long tube of length $L$ oriented along an $x$ axis. The radius of
the tube, $r$, varies with the position along the tube as 
\begin{equation}
r=\frac{r_{0}}{1-a\cos \left( 2\pi x/l\right) }\;,  \label{rvsx}
\end{equation}
where $r_{0}$ is the average radius of the tube, $l$ is the wavelength along
the tube and $a$ is the dimensionless amplitude of the oscillation, see
Fig.\ \ref{fig1}. We assume $L\gg l$. 

We now imagine a bubble in a tube segment with length $l$\ limited by
interfaces at $x_{i}$ and $x_{f}$. The fluid in the bubble (``oil") is 
less wetting with respect to the tube walls than the fluid outside the
bubble (``water").

The capillary pressure drop across the interface between the two
fluids at $x_{i}$ is \cite{b88} 
\begin{equation}
\frac{2\sigma _{wo}}{r(x_{i})}=\frac{2\sigma _{wo}}{r_{0}}\ \left[ 1-a\cos
\left( \frac{2\pi x_{i}}{l}\right) \right] \;,  \label{fxi}
\end{equation}
and across the interface at $x_{f}$, 
\begin{equation}
-\frac{2\sigma _{wo}}{r(x_{f})}=-\frac{2\sigma _{wo}}{r_{0}}\ \left[ 1-a\cos
\left( \frac{2\pi x_{f}}{l}\right) \right] \;.  \label{fxf}
\end{equation}
Here, $\sigma _{wo}$ is the surface tension between the two fluids.
We sum the two capillary pressure drops to get 
\begin{equation}
p_{c}(x_{b})=\frac{4\sigma _{wo}a}{r_{0}}\ \sin \left( \frac{\pi \Delta x_{b}
}{l}\right) \ \sin \left( \frac{2\pi x_{b}}{l}\right) \;,  \label{sxixf}
\end{equation}
where we have defined $x_{b}=(x_{i}+x_{f})/2$ and $\Delta x_{b}=x_{f}-x_{i}$.
We assume the length of the bubble $\Delta x_{b}$ to be smaller than the
wavelength $l$, which is also equal to the length of the tube segment.
Furthermore we have chosen the origin of the $x$-coordinate such the whole
bubble is located between 0 and $l$.

We assume the viscosities of the two fluids to be $\mu_w$ (``water") and
$\mu_o$ (``oil") respectively.  In order to derive a constitutive equation
between volumetric flux $q$ and pressure drop along the tube, 
$\Delta P = (L/l) \Delta p$, we consider the 
entropy production of a tube segment with an average cross-sectional
area $\pi r_{0}^{2}$\ and length $l$ \cite{kbjg10},
\begin{equation}
\frac{dS}{dt}=\int_{0}^{l}dx\ \dot{s}(x)\;,  \label{ep}
\end{equation}
where $\dot{s}(x)$ the entropy
production in the tube per unit of length at position $x$. We assume
the temperature to be constant along the tube.  The entropy
production in the fluid sections times the temperature is equal to 
$-q\partial p(x)/\partial x$, where $q$ is the volumetric flux and $p(x)$ the
pressure at $x$. Because of the incompressible nature of the flow, the 
volumetric flux is independent of position. 
There is no excess entropy production at
the interfaces between the fluid and the bubble, assuming equilibrium
between the two immiscible fluids, cf.\ Eqs.\ (\ref{fxi}) and (\ref{fxf}).
Substituting $\dot{s}(x)$ into Eq.\ (\ref{ep}) and integrating gives 
\begin{eqnarray}
\Theta\frac{dS}{dt} 
&=&-q\left( p_{iw}-p_{1}\right) -q\left( p_{fo}-p_{io}\right)
-q\left( p_{2}-p_{fw}\right)   \nonumber \\
&&  \label{ep2} \\
&=&-q\left( \Delta p-p_{c}(x_{b})\right)   \label{ep2b}
\end{eqnarray}
where $\Theta$ is the temperature, $\Delta p=p_{2}-p_{1}$ 
is the pressure drop across a length $l$ and 
$p_{c}(x_{b})=p_{io}-p_{iw}+p_{fw}-p_{fo}$ is the capillary pressure difference
given in
Eq.\ (\ref{sxixf}). The resulting linear force-flux relations for the pressure
differences in the three sections are written in the form
\begin{eqnarray}
p_{iw}-p_{1} &=&-R_{w}x_{i}q  \nonumber \\
p_{fo}-p_{io} &=&-R_{o}\Delta x_{b}q  \nonumber \\
p_{2}-p_{fw} &=&-R_{w}\left( l-x_{f}\right) q  \label{ff}
\end{eqnarray}
where $R_{w}$ and $R_{o}$\ are the Onsager resistivities per unit of length
for the flow of water and bubble phases, respectively. Using Poisseulle flow
the Onsager resistivities are equal to $8\mu /(\pi r_{0}^{4})$, where $\mu$
is the viscosity of the fluid considered. The resulting pressure
differences are 
\begin{eqnarray}
p_{iw}-p_{1} &=&-\frac{8}{\pi r_{0}^{4}}x_{i}\mu _{w}q  \nonumber \\
p_{fo}-p_{io} &=&-\frac{8}{\pi r_{0}^{4}}\Delta x_{b}\mu _{o}q  \nonumber \\
p_{2}-p_{fw} &=&-\frac{8}{\pi r_{0}^{4}}\left( l-x_{f}\right) \mu _{w}q
\label{ff2}
\end{eqnarray}
The sum of the pressure differences gives the constitutive equation for 
motion of a single bubble 
\begin{equation}
q=-\frac{\pi r_{0}^{4}}{8\mu _{av}l}\ \left( \Delta p-p_{c}(x_{b})\right)\;. 
\label{darcy}
\end{equation}
where 
\begin{equation}
\label{avemu}
\mu _{av}=\frac{1}{l}\ 
\left[\left( l-\Delta x_{b}\right) \mu _{w}+\Delta x_{b}\mu_{o}
\right] 
\end{equation}
is the average viscosity of the two fluids, which we note is
independent of the position of the bubble. One may obtain a similar
expression directly from Eq.\ (\ref{ep2b}).  Eqs.\ (\ref{darcy}) and
(\ref{avemu}) were first derived by Washburn \cite{w21}, but by a 
different route.  The present procedure clarifies 
why Eq.\ (\ref{darcy}) should contain the average viscosity and how the
contribution due to the capillary pressure arises.

The center of mass coordinate of the bubble moves as 
$\pi r_{0}^{2}\dot{x}_{b}=q$. Hence, we have the equation of motion 
\begin{equation}
\dot{x}_{b}=-\frac{r_{0}^{2}}{8l\mu _{av}}\left[ \Delta p-\gamma \sin \left(
\frac{2\pi x_{b}}{l}\right) \right] \;,  \label{eqmotx}
\end{equation}
where we have defined 
\begin{equation}
\gamma =\frac{4\sigma a}{r_{0}}\ \sin \left( \frac{\pi \Delta x_{b}}{l}
\right) \;.  \label{gamma}
\end{equation}
We introduce the angle variable $\theta =2\pi x_{b}/l$ and the time
scale $\tau =\gamma t\pi r_{0}^{2}/(4l^{2}\mu _{av})$. We assume the
pressure drop to be negative, $\Delta p=-|\Delta p|$. The equation of motion
(\ref{gamma}) then becomes 
\begin{equation}
\frac{d\theta }{d\tau }=\frac{|\Delta p|}{\gamma }+\sin \theta \;,
\label{eqmottheta}
\end{equation}
which is nothing but the equation of motion for the overdamped driven
pendulum \cite{s94}.

The period $T_{\tau }$ (measured in the same units as $\tau $), needed for
the bubble to move from one end of the tube with length $l$ to the other end
for a given $\Delta p$, is 
\begin{equation}
T_{\tau }=\int_{0}^{T_{\tau }}d\tau =\int_{0}^{2\pi }\frac{d\theta }{d\theta
/d\tau }=\frac{2\pi \gamma }{\sqrt{\Delta p^{2}-\gamma ^{2}}}\;,
\end{equation}
In seconds this time is
\begin{equation}
T=\frac{4l^{2}\mu _{av}}{\gamma \pi r_{0}^{2}}T_{\tau }=\frac{8l^{2}\mu _{av}
}{r_{0}^{2}\sqrt{\Delta p^{2}-\gamma ^{2}}}\;. \label{tavet}
\end{equation}
We then define the time-averaged angular speed as 
\begin{eqnarray}
\langle \frac{d\theta }{d\tau }\rangle  &=&\frac{1}{T_{\tau }}
\int_{0}^{T_{\tau }}\frac{d\theta }{d\tau }d\tau =\frac{1}{T_{\tau }}
\int_{0}^{2\pi }d\theta   \nonumber \\
&=&\frac{2\pi }{T_{\tau }}=\frac{1}{\gamma }\ \sqrt{\Delta p^{2}-\gamma ^{2}}
\;.  \label{avetheta}
\end{eqnarray}
Now, transforming back from $d\theta /d\tau $ to $\dot{x}_{b}$ to volumetric
flux, $q$, we find the time-averaged flux equation 
\begin{equation}
\langle q\rangle = - \frac{\pi r_0^4}{8\mu_{av}l} \mathrm{sgn}
(\Delta p)\Theta(|\Delta p|-\gamma)
\sqrt{\Delta p^{2}-\gamma ^{2}}\;,  \label{effdarcy}
\end{equation}
where $\mathrm{sgn}$ is the sign function and $\Theta $ is the Heaviside
function. Hence, for $|\Delta p|<\gamma $ there is no flow
through the tube. Furthermore, if there is flow, 
$T\langle\dot{x}_{b}\rangle=l$, as one would expect.

We see that we can write Darcy's law for the time averaged volume flux only
if $\gamma =0$, which is the case if the tube has a constant radius ($a=0$).
The deviation from this law for the time average is due to a capillary
pressure which varies as a function of the position of the bubble along the
tube. The effective flux equation (\ref{effdarcy}) has a threshold pressure
that must be overcome to induce flow, $\gamma $ defined in Eq.\ (\ref{gamma}).
Close to this threshold, when $|\Delta p|-\gamma \ll \gamma $, the
average flow equation becomes 
\begin{equation}
\langle q\rangle =-\frac{\pi r_{0}^{4}}{8\mu _{av}l}\sqrt{2\gamma }\mathrm{
sgn}(\Delta p)\Theta (|\Delta p|-\gamma )\sqrt{|\Delta p|-\gamma }\;,
\label{squareroot}
\end{equation}
i.e., there is 
a square root singularity. As shown in Ref.\ \cite{s94}, this square
root is a consequence of the quadratic extremum of $r$ in Eq.\ (\ref{rvsx})
leading to a saddle-node bifurcation, and therefore it does not depend on the
specific sinusoidal shape of the profile chosen. The smaller the value of 
$r_{0},$ the larger is the threshold value. A radius of 10 micrometer can give
a threshold of the order of 80 bar, which is non-negligible for most
practical purposes.

We may generalize these considerations to a tube segment with length 
$L=\left( 2N+1\right) l$ in which there are $2N+1$ bubbles numbered from $-N$
to $N$. Each bubble $j$ may be characterized by a center of mass position 
$x_{j}$ and a width $\Delta x_{j}$. In view of the incompressible nature of
the flow the volume flux $q$ is independent of the position. This implies
that the velocity $\dot{x_{j}}$ of all the bubbles is the same, $\dot{x_{0}}=
\dot{x_{j}}$.\ The equation of motion (\ref{eqmotx}) may then be generalized
as follows,
\begin{equation}
\dot{x_{0}}=-\frac{r_{0}^{2}}{8L\mu _{av}}\left[ \Delta
P-\sum_{j=-N}^{+N}\gamma _{i}\sin \left( \frac{2\pi }{l}(x_{0}+\delta
x_{j})\right) \right] \;,  \label{eqmotxmult}
\end{equation}
where $\Delta P$ is the total pressure drop over the whole tube segment and 
$\delta x_{j}\equiv x_{j}-x_{0}$. Furthermore we define
\begin{equation}
\gamma _{j}=\frac{4\sigma a}{r_{0}}\ \sin \left( \frac{\pi \Delta x_{j}}{l}
\right) \;.  \label{gammai}
\end{equation}
By using trivial trigonometric identities, we may rewrite this expression as 
\begin{equation}
\dot{x_{0}}=-\frac{r_{0}^{2}}{8L\mu _{av}}\left[ \Delta P-\Gamma _{s}\sin
\left( \frac{2\pi x_{0}}{l}\right) -\Gamma _{c}\cos \left( \frac{2\pi x_{0}}{
l}\right) \right]  \label{trig}
\end{equation}
where 
\begin{equation}
\Gamma _{s}=\sum_{j=-N}^{+N}\gamma _{j}\sin \left( \frac{2\pi \delta x_{j}}{l
}\right) \;,  \label{gammas}
\end{equation}
and 
\begin{equation}
\Gamma _{c}=\sum_{j=-N}^{+N}\gamma _{j}\cos \left( \frac{2\pi \delta x_{j}}{l
}\right) \;.  \label{gammac}
\end{equation}
It should be noted that $\Gamma _{s}$\ and $\Gamma _{c}$\ are
proportional to the number of segments $2N+1$ with length $l$\ and therefore
to the total length $L$.

On non-dimensional form, Eq.\ (\ref{trig}) becomes 
\begin{equation}
\frac{d\theta }{d\tau }=\left( \frac{|\Delta P|}{\Gamma _{s}}\right) +\sin
\theta +\frac{\Gamma _{c}}{\Gamma _{s}}\cos \theta \;.
\label{eqmotthetamult}
\end{equation}
Choosing this form, we have assumed $\Gamma _{s}>\Gamma _{c}$. Hence, the
saddle-node bifurcation will occur in the sine term and not in the cosine
term, which may be ignored. Working through the arguments leading to 
(\ref {effdarcy}), we end up with the same effective flux equation as for the
one-bubble case, but with $\gamma $ substituted for $\Gamma _{s}$. If, on
the other hand, $\Gamma _{c}>\Gamma _{s}$, we may shift $\theta $ by $\pi /2$,
and we are back to Eq.\ (\ref{eqmotthetamult}), but with $\Gamma _{s}$ and 
$\Gamma _{c}$ interchanged. Hence, again we find an effective flux equation
as (\ref{effdarcy}), but with $\gamma $ substituted for $\Gamma _{c}$.

Hansen and Ramstad \cite{hr09} proposed to approach steady-state 
immiscible two-phase flow in porous media using the methods of 
statistical mechanics.  A central quantity in this context is the 
{\it configurational probability\/} $\Pi\{\mathrm{configuration}\}$. 
Possessing this quantity makes it possible to calculate the average of
any quantity associated with the flow.
We now derive the configurational probability for bubbles in a capillary tube.  
The derivation is based on mapping the time average on to a configurational
average.

Time averaging assumes that the states in each time interval are equally
probable. The state of the tube at time $t$ is characterised by the position 
$x_{b}(t)$ of the bubble. The time average of a function $f(x_{b}(t))$ is
therefore given by
\begin{equation}
\langle f\rangle =\frac{1}{T}\int_{0}^{T}f(x_{b}(t))dt\;.
\end{equation}
Using the relation between $x_{b}$\ and $t$\ we may write this average as an
average over $x_{b}$\ in the following way:
\begin{eqnarray}
&\langle f\rangle=\frac{1}{T}\int_{0}^{l}f(x_{b})\frac{1}{dx_{b}/dt}dx_{b}  
\nonumber \\
&=\frac{1}{l}\sqrt{\Delta p^{2}-\gamma ^{2}}\int_{0}^{l}\frac{f(x_{b})}{
\left\vert \Delta p\right\vert +\gamma \sin \left( 2\pi x_{b}/l\right) }
dx_{b}\;.
\end{eqnarray}
This may be written as
\begin{equation}
\langle f\rangle=\int_{0}^{l}\Pi (x_{b})f(x_{b})dx_{b}\;,
\end{equation}
where 
\begin{equation}
\Pi (x_{b})=\frac{1}{T(dx_{b}/dt)}=\frac{\sqrt{\Delta p^{2}-\gamma ^{2}}}{
l\left( \left\vert \Delta p\right\vert +\gamma \sin \left( 2\pi
x_{b}/l\right) \right) }    \label{pi}
\end{equation}
is the probability that the bubble has the position $x_{b}$. This
probability distribution can be interpreted as the probability distribution
of an ensemble of tubes. Hence, it is the {\it configurational 
probability.\/}

The configurational probability depends on the manner
the flow is controlled. If $q$ is kept constant $\Pi (x_{b})=(T\dot{x}
_{b})^{-1}=1/l$. If $\Delta p$\ is kept constant one finds the value given
on the right hand side of Eq.\ (\ref{pi}). Whether $q$\ or $\Delta P$\ is kept
constant is comparable to the choice of an ensemble.

It is interesting to calculate a few averages in the constant $\Delta P$\
ensemble. For the average velocity we find using Eqs.(\ref{tavet}) and 
(\ref{pi}) that
\begin{eqnarray}
\langle \dot{x}_{b}\rangle&=&\int_{0}^{l}\Pi (x_{b})\dot{x}_{b}(x_{b})dx_{b}  
\nonumber \\
&=& \frac{l}{T}
=\frac{
r_{0}^{2}\sqrt{\Delta p^{2}-\gamma ^{2}}}{8l\mu _{av}}\;.
\end{eqnarray}
We calculate the average potential energy stored associated with the 
capillary forces, using Eqs.\ (\ref{darcy}), (\ref{pi})
and $p_{c}(x_{b})=\gamma \sin \left( 2\pi x_{b}/l\right) $, finding
\begin{eqnarray}
\langle p_{c}q\rangle&=&\int_{0}^{l}\Pi (x_{b})p_{c}(x_{b})q(x_{b})dx_{b}  
\nonumber \\
&=&\frac{\pi \gamma r_{0}^{4}}{8\mu _{av}l^{2}}\sqrt{\Delta p^{2}-\gamma ^{2}
}\int_{0}^{l}\sin \left( \frac{2\pi x_{b}}{l}\right) dx_{b} \nonumber\\
&=&0\;.
\end{eqnarray}
The average potential energy associated with the capillary forces is 
therefore zero as it must be.

We proceed to consider an ensemble of single tube segments. For the ensemble
we may interprete $\Pi (x_{b})$\ as the probability that the bubble has the
position $x_{b}$. For the ensemble of tubes this contributes $k_{\text{B}
}\ln l\Pi (x_{b})$\ to the entropy density per unit of length along the
tube. Together with the other entropy contributions in the single tube we
then have 
\begin{eqnarray}
S &=&S^{0}+\pi r_{0}^{2}
\left[ \Delta x_b s_{\text{o}}+\left( l-\Delta x_b\right) s_{\text{w}}
\right]   \nonumber \\
&&-k_{\text{B}}\int_{0}^{l}\Pi(x_{b})\ln l\Pi (x_{b})dx_{b}\;.
\end{eqnarray}
From thermodynamics it follows that the
entropy densities per unit of volume of the single component bulk phases are
given by $s_{\text{o}}=-\partial p_{\text{o}}/\partial \Theta$\ 
and $s_{\text{w}}=-\partial p_{\text{w}}/\partial \Theta$. 
The reference value $S^{0}$\ is due to
other contributions to the entropy, like the excess entropies of the
surfaces, which are constant and which we do not need to consider
explicitly.  The last term, which is the configurational entropy, is constant.

The configurational probability Eq.\ (\ref{pi}) may be generalized to
a network of tubes, i.e., a porous medium \cite{shbk12}.  This makes
it possible to fulfill the program sketched by Hansen and Ramstad
\cite{hr09}.

We have in this note demonstrated that the average flux-pressure relation
in a tube with varying diameter is non-linear and controlled by a square-root
singularity.  We have shown this by deriving the equation of motion from
considering the capillary forces and the entropy production associated with 
the viscous flow.  We have also derived the configurational probability,
i.e., the probability of bubble configurations from the equation of
motion.  From this probability the average of any quantity associated with the
flow may be calculated.

S.\ S.\ thanks the Norwegian Research Council for financial support.




\begin{thebibliography}{9}

\bibitem{d79} F.\ A.\ L\ Dullien, \textsl{Porous Media --- Fluid
Transport and Pore Structure\/} (Academic, New York, 1979).

\bibitem{b88} J.\ Bear, \textsl{Dynamics of Fluids in Porous Media\/}
(Dover Publ.\ Comp., Mineola, 1988).

\bibitem{a92} P.\ M.\ Adler, \textsl{Porous Media: Geometry and 
Transport\/} (Butterworth-Heinemann, Stoneham, 1992).

\bibitem{s95} M.\ Sahimi, \textsl{Flow and Transport in Porous Media\/}
(VCH, Boston, 1995).

\bibitem{gr93} A.\ K.\ Gunstensen and D.\ H.\ Rothman, J.\ Geophys.\ Res.\
{\bf 98}, 6431 (1993).

\bibitem{ap95} D.\ G.\ Avraam and A.\ C.\ Payatakes, J.\ Fluid Mech.\ {\bf 293},
207 (1995); D.\ G.\ Avraam and A.\ C.\ Payatakes, Transp.\ Por.\ Media
{\bf 20}, 135 (1995).

\bibitem{ap99} D.\ G.\ Avraam and A.\ C.\ Payatakes, Ind.\ Eng.\ Chem.\ Res.\
{\bf 38}, 778 (1999).

\bibitem{kah02} H.\ A.\ Knudsen, E.\ Aker and A.\ Hansen, Transp.\ Por.\
Media, {\bf 47}, 99 (2002).

\bibitem{kh02} H.\ A.\ Knudsen and A.\ Hansen, Phys.\ Rev.\ E {\bf 65}, 056310
(2002).

\bibitem{tkrlmtf09} K.\ T.\ Tallakstad, H.\ A.\ Knudsen, T.\
Ramstad, G.\ L{\o}voll, K.\ J.\ M{\aa}l{\o}y, R.\ Toussaint and
E.\ G.\ Flekk{\o}y, Phys.\ Rev.\ Lett.\ {\bf102}, 074502 (2009).

\bibitem{tlkrfm09} K.\ T.\ Tallakstad, G.\ L{\o}voll, H.\ A.\ Knudsen,
T.\ Ramstad, E.\ G.\ Flekk{\o}y and K.\ J.\ M{\aa}l{\o}y, Phys.\ Rev.\ E,
{\bf 80}, 036308 (2009).

\bibitem{rcs11} E.\ M.\ Rassi, S.\ L.\ Codd and J.\ D.\ Seymour, New J.\
Phys.\ {\bf 13}, 015007 (2011).

\bibitem{sh12} S.\ Sinha and A.\ Hansen, Europhys.\ Lett.\ in press
(2012).

\bibitem{c06} M.\ Carlson, {\it Practical Reservoir Simulation\/}
(PennWell Corporation, Tulsa, 2006).

\bibitem{w21} E.\ W.\ Washburn, Phys.\ Rev.\ {\bf 17}, 273 (1921).

\bibitem{kbjg10} S.\ Kjelstrup, D.\ Bedeaux, E.\ Johannessen and J.\ 
Gross, \textsl{Non-Equilibrium Thermodynamics for Engineers\/}
(World Scientific, Singapore, 2010).

\bibitem{s94} S.\ H.\ Strogatz, \textsl{Non-Linear Dynamics and Chaos\/}
(Perseus Press, Cambridge, 1994).

\bibitem{hr09} A.\ Hansen and T.\ Ramstad, Comp.\ Geosci.\ {\bf 13},
227 (2009).

\bibitem{shbk12} S.\ Sinha, A.\ Hansen, D.\ Bedeaux and S.\ Kjelstrup,
in preparation (2012).

\end{thebibliography}
\end{document}